\documentclass[11pt]{article}
\usepackage{amssymb}
\usepackage{colortbl}
\usepackage{amsfonts,amsmath, longtable}

\newcommand{\tr}{{\rm tr}}

\newcommand{\mL}{{\mathcal L}}
\newcommand{\mM}{{\mathcal M}}

\newcommand{\vth}{\vartheta}

\newcommand{\Mat}{ {\rm Mat}(N,\mathbb C) }

\newcommand{\mC}{\mathbb C}

\newcommand{\h}{\hbar}

\def\beq{\begin{equation}}
\def\eq{\end{equation}}
\def\p{\partial}

\def\res{\mathop{\hbox{Res}}\limits}

\begin{document}

\setcounter{page}{1}

\begin{center}


\vspace{0mm}

{\Large{\bf  Classical integrable spin chains }}

\vspace{3mm}

{\Large{\bf
of Landau-Lifshitz type from R-matrix identities
}}



 \vspace{12mm}

 {\Large {D. Domanevsky}}$\,^{\bullet}$
\qquad\quad\quad
 {\Large {A. Zotov}}$\,^{\diamond}$

  \vspace{8mm}


$\bullet$ --
{\em 
Lomonosov Moscow State University, Moscow, 119991, Russia}

$\diamond$ --
 {\em Steklov Mathematical Institute of Russian
Academy of Sciences,\\ 8 Gubkina St., Moscow 119991, Russia}


 {\small\rm {e-mails: danildom09@gmail.com, zotov@mi-ras.ru}}

\end{center}

\vspace{0mm}

\begin{abstract}
We describe a family of 1+1 classical integrable space-discrete models of the Landau-Lifshitz type
through the usage of ansatz for $U$-$V$ (Lax) pair with spectral parameter satisfying the semi-discrete
Zakharov-Shabat equation. The ansatz for $U$-$V$ pair is based on $R$-matrices satisfying the
associative Yang-Baxter equation and certain additional properties.
Equations of motion are obtained using a set of $R$-matrix identities.
In the continuous limit we reproduce the previously known family of the higher rank Landau-Lifshitz equations.
\end{abstract}

%


\parskip 5pt plus 1pt   \jot = 1.5ex




\subsection*{\underline{Introduction}}
Integrable tops of the Euler-Arnold type are integrable models described by the following equations of
motion:
 \beq\label{q01}
 \begin{array}{c}
   \displaystyle{
{\dot S}=[S,J(S)]\,,
 }
 \end{array}
  \eq
  where $S\in\Mat$ is a matrix, which elements $S_{ij}$ are dynamical variables, and
  $J(S)$ is some linear map, that is $J(S)=\sum_{i,j,k,l=1}^N E_{ij}J_{ij,kl}S_{kl}$ with
  certain constants $J_{ij,kl}$, and $E_{ij}$: $(E_{ij})_{ab}=\delta_{ia}\delta_{jb}$ are the standard unit matrices.
  From the classical mechanics viewpoint $J(S)$ is an inverse inertia tensor.

  A wide family of such models was described in \cite{LOZ14,LOZ142,LOZ16} using a special class of quantum
  $R$-matrices. The quantum (non-dynamical) $R$-matrices, by definition, are solutions to the quantum Yang-Baxter
  equation
  \beq\label{q02}
R_{12}^\hbar(z_1-z_2)R_{13}^\hbar(z_1-z_3)R_{23}^\hbar(z_2-z_3)=
R_{23}^\hbar(z_2-z_3)R_{13}^\hbar(z_1-z_3)R_{12}^\hbar(z_1-z_2)\,,
  \eq
 where $\hbar$ is the Planck constant and $z_1,z_2,z_3$ are spectral parameters.
 We assume that $R_{12}^\hbar(z)$ is a $\Mat^{\otimes 2}$-valued function of $z$ and $\hbar$,
 and $R_{ij}^\hbar(z)$ in (\ref{q02}) acts nontrivially on the $i$-th and $j$-th tensor components of vector space
 $\mC^N\otimes \mC^N\otimes \mC^N$.
 The tensor notations are standard, see
 \cite{Skl2,FT}. In this paper we deal with a special class of $R$-matrices, which satisfy not only the equation
 (\ref{q02}) but also\footnote{Two Yang-Baxter equations (\ref{q02}) and (\ref{q03}) have
 different but intersecting sets of solutions. In this paper we impose additional requirements to
 solutions of (\ref{q03}), which make them also be solutions of (\ref{q02}). This
 is why we talk about a subset of solutions of (\ref{q02}) satisfying also (\ref{q03}).} the so-called associative Yang-Baxter equation \cite{FK}:
  \beq\label{q03}
  R^\hbar_{12}(z_{12})
 R^{\eta}_{23}(z_{23})=R^{\eta}_{13}(z_{13})R_{12}^{\hbar-\eta}(z_{12})+
 R^{\eta-\hbar}_{23}(z_{23})R^\hbar_{13}(z_{13})\,,\quad
 z_{ab}=z_a-z_b\,.
  \eq
We consider not all solutions of (\ref{q03}) but those obeying additional properties which are described
below. In particular, an $R$-matrix satisfying (\ref{q03}) is assumed to have the quasi-classical expansion in the
form
  \beq\label{q04}
  \begin{array}{c}
      \displaystyle{
R^\h_{12}(z)=\frac{1}{\h}\,1_N\otimes 1_N+r_{12}(z)+\h\,
m_{12}(z)+O(\h^2)\in\Mat^{\otimes 2} }\,,
  \end{array}
  \eq
where $1_N$ is the identity $N\times N$ matrix, $r_{12}(z)$ is the classical $r$-matrix,
and the next coefficient $m_{12}(z)$ provides
explicit expression for the inverse inertia tensor in $J(S)$ in the equation (\ref{q01}) of classical
Hamiltonian mechanics:
  \beq\label{q05}
  \begin{array}{c}
      \displaystyle{
J(S)=\tr_2(m_{12}(0)S_2)\,,\qquad S_2=1_N\otimes S\,,
}
  \end{array}
  \eq
where $\tr_2$ is the trace over the second tensor component.
 The top with $J(S)$ (\ref{q05}) can be generalized to the ''relativistic'' top \cite{LOZ14,LOZ142,KrZ} depending on
 the deformed linear map $J^\eta(S)$. This is similar to how the Ruijsenaars-Schneider model
 generalizes the Calogero-Moser system.
 The deformation parameter $\eta$ plays the role of the Planck constant in some $R$-matrix $R_{12}^\eta(z)$.
 A review of the integrable systems of this type including possible generalizations and applications can be found in
 \cite{GSZ,KrZ}.

 The field generalization of the integrable top (\ref{q01}) to 1+1 integrable field theory can be performed in different ways. We mention only two approaches, which we use in this paper.
  The first one is the most fundamental and widely known one \cite{Skl,Skl2,FT}. It is based
 on the classical quadratic $r$-matrix structure of the Sklyanin type, providing integrable spin chains. The field
 theory of the Landau-Lifshitz type \cite{LL} arises in the continuous limit. The second approach  deals
 with a special ansatz for $U$-$V$ pairs for the Zakharov-Shabat (zero curvature) equation
  \beq\label{q06}
  \begin{array}{l}
  \displaystyle{
 \p_t U(z)-\p_x V(z)+[U(z),V(z)]=0\,,\qquad U(z),V(z)\in\Mat\,.
 }
 \end{array}
 \eq
 This approach uses jointly
 2d generalization of Hitchin systems \cite{LOZ} and the construction based on the associative Yang-Baxter equation
 \cite{AtZ2,DLOZ}. In particular, it was shown in \cite{AtZ2} that the integrable top (\ref{q01}) with
 $J(S)$ (\ref{q05}) is generalized to 1+1 field theory of the Landau-Lifshitz type with equations of motion
 \beq\label{q07}
  \begin{array}{c}
  \displaystyle{
  \p_t S=\frac{1}{c}\,[S,\p^2_x S]+\frac{2c}{N}\, [S,J(S)]-2[S,E^0(\p_x S)]\,,
 }
 \end{array}
 \eq
where $S=S(t,x)\in\Mat$ is a matrix\footnote{
Equations of motion in the form (\ref{q07}) arise in the special case when
the matrix $S$ is of rank one, see (\ref{q25}) below and the comment after it.
} of dynamical field variables, $c$ is some constant and $E^0$ is another one linear
map. It will be defined below. Here we mention that it vanishes in the $N=2$ case thus providing the standard
Landau-Lifshitz equation \cite{LL,Skl} for the vector $\vec{S}(t,x)=(S_1,S_2,S_3)$, which components are the components
of the traceless part of matrix $S$ in the Pauli matrices basis. Recent results on the field generalizations
of finite-dimensional integrable systems can be found in \cite{ZZ,AtZ2,Z24,DLOZ}.


\noindent {\bf Purpose of the paper} is to fill the lower right corner on the following scheme:
  \beq\label{q08}
  \begin{array}{ccc}
\underline{\hbox{Integrable top}} & \stackrel{\hbox{\footnotesize{non-relat. limit}}}{\longleftarrow} &
\underline{\hbox{Relativistic integrable top}}
\\
\downarrow\hbox{\footnotesize{2d version}} &  & \downarrow\hbox{\footnotesize{2d version}}
\\
\underline{\hbox{Landau-Lifshitz model}} & \stackrel{\hbox{\footnotesize{continuous limit}}}{\longleftarrow}
& \underline{\hbox{Space-discrete L-L model}}
 \end{array}
 \eq
That is we describe the discrete version of the higher rank Landau-Lifshitz equation (\ref{q07}) from \cite{AtZ2}.
For this purpose we use the standard construction of classical spin chains \cite{Skl2,FT} and combine it
with the description of the relativistic top through $R$-matrices \cite{LOZ14,KrZ}. In fact,
this method was used in \cite{ZZ}, where the elliptic spin chain of this type was described.
We formulate similar result for an arbitrary $R$-matrix satisfying (\ref{q03}) together with some additional properties.
%
In particular, our construction is valid for elliptic ${\rm GL}_N$ Baxter-Belavin $R$-matrix and
different type trigonometric and rational degenerations including 7-vertex trigonometric and 11-vertex rational $R$-matrices.
These type $R$-matrices were studied in \cite{Chered,KrZ}.
%
Then we show that the obtained equations reproduce (\ref{q07}) in the continuous limit. Let us also mention the
paper \cite{Nijhoff}, where the fully discrete version of the elliptic Landau-Lifshitz model was described. From viewpoint
of our approach that model is rather the fully discrete version of the Ruijsenaars-Schneider model, see \cite{ZZ}.
Different type discretizations of the soliton equations of Landau-Lifshitz type are also known from \cite{Adler}.

\subsection*{\underline{Integrable tops from $R$-matrix identities}}
We begin with a brief description of integrable tops based
on the $R$-matrix identities \cite{LOZ14,LOZ16,KrZ,GSZ}.
\paragraph{$R$-matrix properties.} Let us first formulate the above mentioned additional properties of the $R$-matrices.
Besides the associative Yang-Baxter equation (\ref{q03}) the $R$-matrices under consideration  satisfy the skew-symmetry\footnote{$P_{12}$
 is the matrix permutation operator. For any pair of matrices $A,B\in\Mat$: $(A\otimes B) P_{12}=P_{12}(B\otimes A)$.}
  \beq\label{q09}
  \begin{array}{c}
  \displaystyle{
 R^\hbar_{12}(z)=-R_{21}^{-\hbar}(-z)=-P_{12}R_{12}^{-\hbar}(-z)P_{12}
 }
 \end{array}
 \eq
and unitarity
 \beq\label{q10}
   \begin{array}{c}
 \displaystyle{
R^\hbar_{12}(z) R^\hbar_{21}(-z) = \phi(\hbar,z)\phi(\hbar,-z)\,\,1_N\otimes 1_N\,,
 }
  \end{array}
  \eq
  where $\phi(\hbar,z)$ is the scalar solution to (\ref{q03}) given by the Kronecker function
 \beq\label{q11}
   \begin{array}{c}
 \displaystyle{
\phi(\hbar,z)=\frac{\vth'(0)\vth(\hbar+z)}{\vth(\hbar)\vth(z)}
\stackrel{\hbox{\footnotesize{trig. limit}}}{\longrightarrow}
\frac{\sin(\pi(\hbar+z))}{\sin(\pi\hbar)\sin(\pi z)}
\stackrel{\hbox{\footnotesize{rat. limit}}}{\longrightarrow}
\frac{\hbar+z}{\hbar z}
 }
  \end{array}
  \eq
  chosen for elliptic, trigonometric or a rational $R$-matrix respectively.
$R$-matrices have only simple poles at $\hbar=0$ and $z=0$ with the residues
  \beq\label{q12}
  \begin{array}{c}
  \displaystyle{
 \res\limits_{\hbar=0}R^\hbar_{12}(z)=1_N\otimes
1_N=1_{N^2}\,,\quad\quad \res\limits_{z=0}R^\hbar_{12}(z)=P_{12}\,.
 }
 \end{array}
 \eq
The local behaviour near $\hbar=0$ is given by the quasi-classical limit (\ref{q04})
and
near $z=0$ we have
  \beq\label{q13}
  \begin{array}{c}
      \displaystyle{
R^\h_{12}(z)=\frac{1}{z}\,P_{12}+R^{\h,(0)}_{12}+zR^{\h,(1)}_{12}+O(z^2)\,,
}
  \end{array}
  \eq
  \beq\label{q14}
  \begin{array}{c}
      \displaystyle{
 R^{\h,(0)}_{12}=\frac{1}{\h}\,1_N\otimes
 1_N+r^{(0)}_{12}+\hbar\, m_{12}(0)+O(\hbar^2)\,,\quad
 r_{12}(z)=\frac{1}{z}\,P_{12}+r^{(0)}_{12}+zr^{(1)}_{12}+O(z^2)\,.
 }
  \end{array}
  \eq
From the skew-symmetry (\ref{q09}) we conclude that
  \beq\label{q15}
  \begin{array}{c}
  r_{12}(z)=-r_{21}(-z)\,,\quad
  m_{12}(z)=m_{21}(-z)\,,\quad
      R^{\h,(0)}_{12}=-R^{-\h,(0)}_{21}\,,\quad
  r_{12}^{(0)}=-r_{21}^{(0)}\,.
  \end{array}
  \eq
If the Fourier symmetry $R^\hbar_{12}(z)P_{12}=R^{z}_{12}(\hbar)$ holds true
then also
  \beq\label{q16}
  \begin{array}{c}
  R^{z,(0)}_{12}=r_{12}(z)P_{12}\,,\quad
      R^{z,(1)}_{12}=m_{12}(z)P_{12}\,,\quad
      r^{(1)}_{12}=m_{12}(0)P_{12}\,,\quad
  r_{12}^{(0)}=r_{12}^{(0)}P_{12}\,.
  \end{array}
  \eq
In what follows
we use the following degeneration of the relation (\ref{q03}), which was also used in \cite{KrZ,GSZ}:
  \beq\label{q161}
  \begin{array}{c}
     \displaystyle{
  R_{13}^\eta(z)r_{12}(z)=R_{12}^\eta(z)R_{23}^{\eta,(0)}-r_{23}^{(0)}R_{13}^\eta(z)
  -P_{23}\p_z R_{13}^\eta(z)+\p_\eta R_{13}^\eta(z)\,,
 }
  \end{array}
  \eq
  \beq\label{q162}
  \begin{array}{c}
     \displaystyle{
  r_{12}(z)R_{13}^\eta(z)=R_{23}^{\eta,(0)}R_{12}^\eta(z)-R_{13}^\eta(z)r_{23}^{(0)}
  -\p_z R_{13}^\eta(z)P_{23}+\p_\eta R_{13}^\eta(z)\,.
 }
  \end{array}
  \eq
\paragraph{Lax pairs.}
In the finite-dimensional mechanics the integrability comes from
the Lax equation
  \beq\label{q17}
  \displaystyle{
  {\dot L}(z)=[L(z),M(z)]\,,\qquad L(z),M(z)\in\Mat\,.
  }
  \eq
The Lax pair for the equation (\ref{q01}) of non-relativistic top
 with $J(S)$ (\ref{q05}) is written in terms of coefficients
of the expansion (\ref{q04}):
%
  \beq\label{q18}
  \begin{array}{c}
     \displaystyle{
 L(z,S)=\tr_2(r_{12}(z)S_2)=\sum\limits_{i,j,k,l=1}^N r_{ij,kl}(z)S_{lk}E_{ij}\,,\quad
  M(z,S)=\sum\limits_{i,j,k,l=1}^N m_{ij,kl}(z)S_{lk}E_{ij}\,.
 }
  \end{array}
  \eq
%
where the explicit expressions are written through
the classical $r$-matrix
 $r_{12}(z)=\sum_{i,j,k,l=1}^N r_{ij,kl}(z) E_{ij}\otimes E_{kl}$
and similarly for $m_{12}(z)$.
For the relativistic top the equations of motion are of the same form as (\ref{q01}) but with the $\eta$-deformed
inverse
inertia tensor $J^\eta(S)$:
  \beq\label{q20}
  \begin{array}{c}
     \displaystyle{
 J^{\eta}(S)=\tr_2\Big(\Big(R^{\eta,(0)}_{12}-r_{12}^{(0)}\Big)S_2\Big)\stackrel{(\ref{q14})}{=}
 \frac{\tr(S)}{\eta}\,1_N+\eta\,\tr_2\Big(m_{12}(0)S_2\Big)+O(\eta^2)
 }
  \end{array}
  \eq
It is written in terms of $R^{\eta,(0)}_{12}$ from (\ref{q13}) and $r_{12}^{(0)}$ from (\ref{q14}).
The Lax pair in this case has the form
  \beq\label{q21}
  \begin{array}{c}
     \displaystyle{
\mL(z,S)=\tr_2(R^\eta_{12}(z)S_2)\,,\qquad
  \mM(z,S)=-\tr_2(r_{12}(z)S_2)\,.
 }
  \end{array}
  \eq
It is interesting to notice that in the non-relativistic limit the roles of $L$ and $M$ matrices get interchanged.
Indeed, when $\eta\rightarrow 0$ we have $\mL(z,S)=\eta^{-1}\tr(S)1_N+L(z,S)+\eta M(z,S)+O(\eta^2)$,
while $\mM(z,S)=-L(z,S)$.

\subsection*{\underline{Higher rank Landau-Lifshitz equations from $R$-matrices}}
In the field theory case the dynamical variables become the fields $S=S(t,x)$.
For definiteness we assume the periodic boundary conditions, that is the space variable $x$
is a coordinate on a unit circle, and $S(t,x+2\pi)=S(t,x)$. According to the general construction
for 1+1 field generalizations of the integrable finite-dimensional systems \cite{LOZ}
the $U$-matrix in the Zakharov-Shabat equation (\ref{q06}) has the same form as in the
finite-dimensional case (this is true for the top-like models under consideration):
  \beq\label{q22}
  \begin{array}{c}
     \displaystyle{
 U(z,S)=\tr_2(r_{12}(z)S_2)\,.
 }
  \end{array}
  \eq
The matrix $V$ is more complicated, see details in \cite{AtZ2}. As a result, one obtains
the equation (\ref{q07}) with $J(S)$ from (\ref{q05}) and the linear map $E^0$ defined as
  \beq\label{q23}
  \begin{array}{c}
     \displaystyle{
 E^0(B)=\tr_2\Big(r_{12}^{(0)}B_2\Big)\,,\quad \forall B\in\Mat\,.
 }
  \end{array}
  \eq
The meaning of the constant $c$ in (\ref{q07}) is as follows. In the finite-dimensional case the
matrix $S$ in (\ref{q01}) is arbitrary. In the field theory case the construction of $U$-$V$ pairs
for (\ref{q06}) requires additional restriction:
  \beq\label{q24}
  \begin{array}{c}
     \displaystyle{
 S^2=c S\,,
 }
  \end{array}
  \eq
that is the eigenvalues of the matrix $S$
are equal to either $c$ or $0$. The corresponding equations of motion were derived in \cite{AtZ2}.
These equations are simplified to (\ref{q07}) in the special case when there is a single eigenvalue which equals $c$
 (and the rest of eigenvalues equal zero). This case corresponds to the rank one matrix:
  \beq\label{q25}
  \begin{array}{c}
     \displaystyle{
 S=\xi\otimes\psi\,,\qquad (\psi,\xi)=c\,,
 }
  \end{array}
  \eq
  where $\xi$ is a $N$-dimensional column-vector, and $\psi$ is a $N$-dimensional row-vector,
  and $(\psi,\xi)$ is their scalar product.
  In fact, integrability of the model (\ref{q22}) holds true for generic matrix $S$ since the
  classical $r$-matrix structure is independent of eigenvalues of $S$, and existence of $r$-matrix
  structure guaranties Poisson commutativity of the traces of powers of the monodromy matrices.
  However, in order to write down explicit equations of motion and explicit $V$-matrix one should
  deal with special matrices of types (\ref{q24}) or (\ref{q25}).
  In what follows we deal with the case (\ref{q25}), i.e. $S_{ij}=\xi_i\psi_j$.

\subsection*{\underline{Space-discrete Landau-Lifshitz equations}}
\paragraph{Spin chain.} In our construction of a periodic chain
we follow \cite{Skl2,FT} but use the relativistic top (\ref{q21}) as a building block.
It was shown in \cite{KrZ} that $\mL(z,S)$ satisfies the quadratic classical $r$-matrix structure
  \beq\label{q26}
  \begin{array}{c}
     \displaystyle{
\{\mL^\eta_1(z,S),\mL^\eta_2(w,S)\}=[\mL^\eta_1(z,S)\mL^\eta_2(w,S),r_{12}(z-w)]\,,
 }
  \end{array}
  \eq
which yields the classical Sklyanin type Poisson brackets
  \beq\label{q27}
  \begin{array}{c}
     \displaystyle{
\{S_1,S_2\}=[S_1
S_2,r_{12}^{(0)}]+[E^\eta(S)_1S_2,P_{12}]\,,\qquad
E^\eta(S)=\tr_3(R_{13}^{\eta,(0)}S_3)\,.
 }
  \end{array}
  \eq
Consider $n$ sites on a unit circle and assign to each site the rank one (\ref{q25}) dynamical matrix
$S^k=\xi^k\otimes\psi^k$, $k=1,..,n$. Let the Poisson brackets be of the form (\ref{q27}) at each site and
$\{S^k_1,S^j_2\}=0$ for $k\neq j$. Then the monodromy matrix
$T(z)=\mL(z,S^1)\mL(z,S^2)...\mL(z,S^n)$ also satisfies (\ref{q26}) thus providing an integrable system,
since it follows from (\ref{q26}) for $T(z)$ that $\{\tr(T(z)),\tr(T(w))\}=0$. Details of this
construction can be found in \cite{ZZ} in the elliptic case.

Main result is the following statement. Introduce notations
  \beq\label{q28}
  \begin{array}{c}
     \displaystyle{
L^k(z)=\mL(z,S^k)=\tr_2\Big(R^\eta_{12}(z)S^k_2\Big)\,,\quad
  M^k(z)=-\tr_2\Big(r_{12}(z)S^{k+1,k}_2\Big)\,,\quad S^{k+1,k}=\frac{\xi^{k+1}\otimes\psi^k}{(\psi^k,\xi^{k+1})}\,.
 }
  \end{array}
  \eq
Then the discrete Zakharov-Shabat equation
  \beq\label{q29}
  \begin{array}{c}
     \displaystyle{
{\dot L}^k(z)-L^k(z)M^k(z)+M^{k-1}(z)L^k(z)=0
 }
  \end{array}
  \eq
holds true identically in $z$ and provides the following equations of motion:
  \beq\label{q30}
  \begin{array}{c}
     \displaystyle{
{\dot S}^k=E^0(S^{k,k-1})S^k-S^kE^0(S^{k+1,k})+S^{k,k-1}E^\eta(S^k)-E^\eta(S^k)S^{k+1,k}\,.
 }
  \end{array}
  \eq
with the notations $E^0$ from (\ref{q23}) and $E^\eta$ from (\ref{q27}). The proof is by direct calculation.
For example, for the term $L^k(z)M^k(z)$ we have
  \beq\label{q31}
  \begin{array}{c}
     \displaystyle{
L^k(z)M^k(z)=\tr_{2,3}\Big(R_{12}^\eta(z)r_{13}(z)S_2^kS_3^{k+1,k}\Big)\,.
 }
  \end{array}
  \eq
Then one should use the $R$-matrix identity (\ref{q161}).
Similarly, $M^{k-1}(z)L^k(z)$ is written through (\ref{q162}).
It is also important to take into account (\ref{q24}) and (\ref{q25}), which also assume
$S^k{S}^{k+1,k}={S}^{k,k-1}S^k=S^k$ and $\tr(S^{k+1,k})=\tr(S^{k,k-1})=N$. In this way the statement
follows\footnote{After calculations one gets expression of the form $\tr_2(R^\eta_{12}(z)(*)_2)=0$,
where $*$ is the l.h.s. of (\ref{q30}) minus the r.h.s. of (\ref{q30}). To see that $*=0$
one should compute $\res\limits_{z=0}\tr_2(R^\eta_{12}(z)(*)_2)=\tr_2(P_{12}(*)_2)=*=0$.}.

\paragraph{1+1 field theory.} The field analogue of the equations (\ref{q30})
is obtained straightforwardly. In the field case the matrices $L^k(z)$, $M^k(z)$ are replaced
with $U(z,x)$ and $V(z,x)$, and the matrix $M^{k-1}(z)$ transforms into $V(z,x-\eta)$. Then
the equation (\ref{q29})  takes the form of the semi-discrete Zakharov-Shabat equation:
  \beq\label{q32}
  \begin{array}{c}
     \displaystyle{
{\dot U}(z,x)-U(z,x)V(z,x)+V(z,x-\eta)U(z,x)=0\,.
 }
  \end{array}
  \eq
It follows from the upper statement that (\ref{q32}) holds true identically in $z$ for
the $U$-$V$ pair (\ref{q28}) written in the field case as
  \beq\label{q33}
  \begin{array}{c}
     \displaystyle{
U(z,x)=\tr_2\Big(R^\eta_{12}(z)S_2(x)\Big)\,,\qquad
  V(z,x)=-\frac{\tr_2\Big(r_{12}(z)\Big(\xi(x+\eta)\otimes\psi(x)\Big)_2\Big)}{(\psi(x),\xi(x+\eta))}\,,
 }
  \end{array}
  \eq
where $S(x)=\xi(x)\otimes\psi(x)$, and $(\psi(x),\xi(x))=\tr(S(x))=c$.
The corresponding equations are obtained from (\ref{q30}) by the substitution
$\xi^k\rightarrow \xi(x)$, $\psi^k\rightarrow \psi(x)$ and
$\xi^{k\pm 1}\rightarrow \xi(x\pm\eta)$, $\psi^{k\pm 1}\rightarrow \psi(x\pm\eta)$.

\subsection*{\underline{Continuous limit}}
Let us show that the defined above discrete model reproduces the Landau-Lifshitz
equation (\ref{q07}) in the continuous limit $\eta\rightarrow 0$.
Using expansions (\ref{q04}), (\ref{q13}), (\ref{q14}) and the Taylor expansion
$\xi(x\pm\eta) = \xi(x) \pm\eta \partial_{x}\xi(x) + \frac{1}{2}\eta^{2}\partial_{x}^{2}\xi(x) + O(\eta^{3})$
for the r.h.s. of the equations of motion (\ref{q30}) one obtains
  \beq\label{q34}
  \begin{array}{c}
     \displaystyle{
-S_{x}+\frac{\eta}{2c}\Big( [S,S_{xx}]+2[S,J(S)] + 2[S,E^0(S_{x})] \Big)+O(\eta^2)\,.
 }
  \end{array}
  \eq
For example, the expression $[S,E^0(S_{x})]$ comes as
  \beq\label{q351}
  \begin{array}{c}
     \displaystyle{
 [S,E^0(S_{x})] = (\xi\otimes\psi)\tr_{2}(r_{12}^{(0)}(\xi_{x}\otimes\psi)_{2}) + (\xi\otimes\psi)\tr_{2}(r_{12}^{(0)}(\xi\otimes\psi_{x})_{2})  -
 }
 \\
    \displaystyle{
 - \tr_{2}(r_{12}^{(0)}(\xi_{x}\otimes\psi)_{2})(\xi\otimes\psi) - \tr_{2}(r^{(0)}_{12}(\xi\otimes\psi_{x})_{2})(\xi\otimes\psi)\,.
 }
  \end{array}
  \eq
To prove it one should also use (\ref{q16}) and a set of identities (see \cite{AtZ2,DLOZ})
  \beq\label{q35}
  \begin{array}{c}
     \displaystyle{
(\xi\otimes\psi)\tr_{2}(r^{(0)}_{12}(\xi\otimes\psi)_{2}) = 0
=(\xi\otimes\psi)\tr_{2}(r^{(0)}_{12}(\xi_{x}\otimes\psi)_{2})\,,
 }
 \\
 \displaystyle{
 \tr_{2}(r^{(0)}_{12}(\xi\otimes\psi)_{2})(\xi_{x}\otimes\psi) =
 \tr_{2}(r^{(0)}_{12}(\xi_{x}\otimes\psi)_{2})(\xi\otimes\psi)\,,
 }
  \\
 \displaystyle{
(\xi\otimes\psi_{x})\tr_{2}(r^{0}_{12}(\xi\otimes\psi)_{2}) =
 -(\xi\otimes\psi)\tr_{2}(r^{(0)}_{12}(\xi\otimes\psi_{x}))\,.
 }
  \end{array}
  \eq
The expression (\ref{q34}) means that we obtain a linear combination of different flows in the continuous model.
By representing ${\dot S}$ in the l.h.s. of (\ref{q30}) as $\p_{t_1}S+\eta\p_{t_2}S+O(\eta^2)$ we get
$\p_{t_1}S=-\p_x S$ and
  \beq\label{q36}
  \begin{array}{c}
     \displaystyle{
\p_{t_2}S=\frac{1}{2c}\Big( 2[S,J(S)] + [S,S_{xx}] + 2[S,E^0(S_{x})] \Big)\,,
 }
  \end{array}
  \eq
which is the Landau-Lifshitz model (\ref{q07}) from \cite{AtZ2} up to some simple redefinitions
(namely, $t_2\rightarrow 2c^2t_2$, $x\rightarrow -cx$ and $J(S)\rightarrow NJ(S)$).

%


\paragraph{Acknowledgments.}



The work of A. Zotov was performed at the Steklov International Mathematical Center and supported by the Ministry of Science and Higher Education of the Russian Federation (agreement no. 075-15-2025-303).
The work of D. Domanevsky was supported by the Foundation
for the Advancement of Theoretical Physics and Mathematics ''BASIS''.


\begin{footnotesize}

\end{footnotesize}

\end{document}